\documentclass[aps,nofootinbib,superscriptaddress,twocolumn,showpacs]{revtex4}
\usepackage{amsmath}
\usepackage{amssymb}
\usepackage{graphicx}
\usepackage{color}

\begin{document}

\title{Comment on ``Novel Attractive Force between Ions in Quantum Plasmas''}
\author{Yu.~Tyshetskiy}
\email{y.tyshetskiy@physics.usyd.edu.au}\
\affiliation{School of Physics, The University of Sydney, NSW 2006, Australia}
\author{S.V.~Vladimirov}
\affiliation{School of Physics, The University of Sydney, NSW 2006, Australia}
\affiliation{Metamaterials Laboratory, National Research University of Information Technology, Mechanics, and Optics, St Petersburg 199034, Russia}

\date{\today}
\received{}

\maketitle


\newpage

In a recent Letter~\cite{SE_PRL_2012}, Shukla and Eliasson reported a new attractive force between ions in degenerate quantum plasmas. This interesting result has potentially significant implications, as it can lead to formation of ion clusters in quantum plasmas, even at densities a few times below the solid density~\cite{SE_clustering}. The suggested important consequence is the modification of the ignition criterion in the fast ignition schemes in inertial confinement fusion.

It is not surprising that this result has already prompted a lively discussion in the literature~\cite{Bonitz_etal_2012,Shukla_etal_response}. One of the reasons is that this prediction is a theoretical one, with no experimental verification done so far. Moreover, the generalized quantum hydrodynamical model used to obtain this result ignores some physics that, when taken into account, may significantly affect the predicted attractive force. Namely, this model, as any hydrodynamical model, completely misses kinetic effects, which in case of degenerate electrons are known to lead to the Kohn singularity in the dielectric response function, which in turn leads to Friedel oscillations in the potential of a charge screened by degenerate electrons~\cite{LL_10,Else_etal_2010}. The depth of (at least some of) the potential wells associated with the Friedel oscillations could be comparable or even exceed the maximum depth of the potential well obtained within the generalized hydrodynamical model of Ref.~\cite{SE_PRL_2012}, as the latter is rather shallow. It is thus important to compare the potentials produced by both models to establish whether and when the predicted attractive part of the Shukla-Eliasson's (SE) screened ion potential is important relative to the attractive parts of such ``Kohn'' potential predicted by the kinetic model.

\begin{figure}
\includegraphics[width=2.8in]{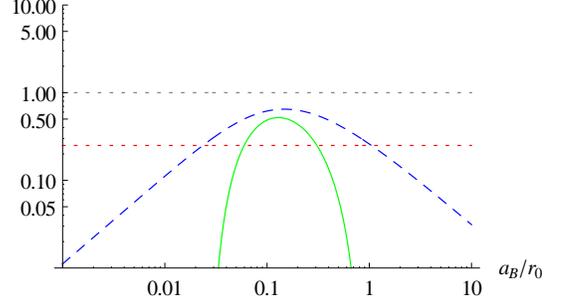}
\caption{\label{fig:eta} (Color online) The attractive potential well depth ratio $\eta$ (\ref{eq:eta}) (solid green line) and the coupling parameters $\alpha$ (dashed blue line) as functions of $a_B/r_0$. The critical value $\alpha=1/4$ is indicated with a dotted red line, and the dotted gray line marks the unity, for visual reference.}
\end{figure}

The generalized quantum hydrodynamical model (GQHD model) used in Ref.~\cite{SE_PRL_2012} yields
a potential $\phi^{\rm SE}$ whose shape depends on the coupling parameter $\alpha$ of the electron gas with exchange interaction and correlations taken into account~\cite{SE_PRL_2012}:
\begin{eqnarray}
\alpha \approx 9.3\pi\frac{a_B}{r_0}\left[1+\left(3\pi^2\right)^{2/3}\frac{a_B}{r_0} + \frac{0.62}{1+18.36\ a_B/r_0}\right]^{-2}, \label{eq:alpha}
\end{eqnarray}
where $a_B=\hbar^2/e^2m$ is the Bohr radius, and $r_0=n_0^{-1/3}$ is the average inter-electron distance in the plasma. The most interesting case is $\alpha>1/4$, when the resulting potential $\phi^{\rm SE}$ has an attractive part. This happens in a limited range of $a_B/r_0$ (see Fig.~\ref{fig:eta}). For $\alpha>1/4$, $\phi^{\rm SE}$ has the following dependence on the dimensionless distance $R=r\sqrt{3}\omega_{pe}/v_F$ from the test charge (here $\omega_{pe}$ is the electron plasma frequency, $v_F=\hbar\left(3\pi^2\right)^{1/3}/m r_0$ is the electron Fermi velocity):
\begin{eqnarray}
\phi^{\rm SE}_{\alpha>1/4} = \frac{\sqrt{3}\omega_{pe}}{v_F}\frac{Q}{R}\left[\cos\left(A_1 R\right)+b \sin\left(A_1 R\right) \right] e^{-A_2 R},   \label{eq:phi_SE}
\end{eqnarray}
with $A_{1,2} = \left({\alpha}/{\alpha_0}\right)^{1/4}\left({\sqrt{4\alpha}\mp1}/{4\alpha}\right)^{1/2}$, $b={1}/{\sqrt{4\alpha-1}}$, and $\alpha_0$ is the coupling parameter of an electron gas with exchange interactions and correlations neglected (ideal electron gas), $\alpha_0 = \left(9/\pi^5\right)^{1/3}\left(r_0/a_B\right)$.
The attractive part of $\phi^{\rm SE}_{\alpha>1/4}$ is most pronounced when the coupling parameter $\alpha$ is at its maximum, $\alpha\approx0.627$ at $a_B/r_0\approx0.15$, corresponding to $n_0\approx2\times10^{22}$~cm$^{-3}$, a few times below solid densities~\cite{SE_clustering}.

The quantum kinetic model (Wigner model)~\cite{LL_10,Klim_Statfizika,Vlad_Tysh_UFN_2011,Tyshetskiy_etal_PoP_2011} accounts for (i) electron degeneracy (via the explicit form of Fermi-Dirac distribution of electrons at equilibrium), (ii) quantum recoil (explicitly included in the Wigner equation), and (iii) kinetic effects arising from the shape of the degenerate electron distribution function (e.g., Kohn singularity in the static response function~\cite{LL_10,Vladimirov_Kohn_1994}). Yet this model, in its simplest form, ignores electron exchange interaction and correlations.
The potential $\phi^{\rm Kohn}$ of a test charge $Q$, yielded by this model is 
\begin{eqnarray}
\phi^{\rm Kohn}(R) &=& \frac{\sqrt{3}\omega_{pe}}{v_F}\frac{2Q}{\pi R}{\rm Im}\int_0^\infty{\frac{e^{iKR}}{\varepsilon(0,K)} \frac{dK}{K}}, \label{eq:phi_Kohn} \\
\varepsilon(0,K) &=& 1 + \frac{1}{2K^2}\left\{1-g_+ + g_-\right\}, \\
g_\pm &=& \frac{\alpha_0 K^2-3}{4\sqrt{3\alpha_0}K}\ln\left(\frac{\sqrt{\alpha_0}K\pm\sqrt{3}}{\sqrt{\alpha_0}K\mp\sqrt{3}}\right),
\end{eqnarray}
where $K=kv_F/\sqrt{3}\omega_{pe}$. Due to the Kohn singularity of $\varepsilon(0,K)$ at $K=\sqrt{3/\alpha_0}$, the potential (\ref{eq:phi_Kohn}) has an oscillatory structure at sufficiently large $R$, i.e., far from the test charge (Friedel oscillations)~\cite{LL_10}.

\begin{figure}
\includegraphics[width=2.8in]{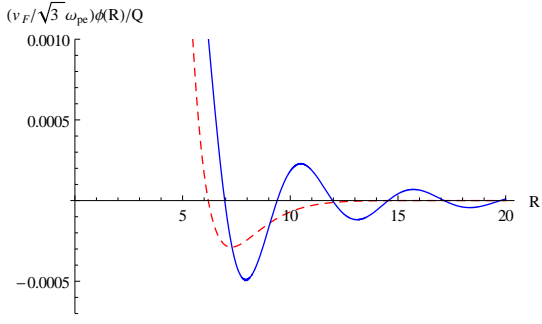}
\caption{\label{fig:phi_comparison} (Color online) The normalized potential $(v_F/\sqrt{3}\omega_{pe})\phi/Q$ as a function of the normalized distance $R=r\sqrt{3}\omega_{pe}/v_F$ from the test charge $Q$, obtained from the GQHD model [Eq.~(\ref{eq:phi_SE}), dashed curve] and from the kinetic model [Eq.~(\ref{eq:phi_Kohn}), solid curve]. The potentials are plotted for $a_B/r_0\approx0.15$, at which the coupling parameter (\ref{eq:alpha}) is maximum, and the attractive part of $\phi^{\rm SE}_{\alpha>1/4}$ is the deepest.}
\end{figure}

Both potentials (\ref{eq:phi_SE}) and (\ref{eq:phi_Kohn}) have attractive parts: $\phi^{\rm SE}_{\alpha>1/4}(R)$ has a single potential well, while $\phi^{\rm Kohn}(R)$ has a series of potential wells associated with Friedel oscillations (see Fig.~\ref{fig:phi_comparison}). The reasons behind these wells are, however, different. The well in $\phi^{\rm SE}_{\alpha>1/4}(R)$ occurs as a result of the interplay between quantum statistical pressure and quantum recoil on one hand, and exchange interaction and correlations of electrons on the other hand; it vanishes if the latter are neglected. The wells in $\phi^{\rm Kohn}(R)$ occur due to the quantum kinetic effects that are neglected in the GQHD model. A comparison of the respective attractive well depths of the potentials (\ref{eq:phi_SE}) and (\ref{eq:phi_Kohn}) is shown in Fig.~\ref{fig:phi_comparison} for the case of maximum $\alpha$ ($a_B/r_0\approx0.15$), when the attractive well of $\phi^{\rm SE}_{\alpha>1/4}(R)$ is the deepest.

Let us define a parameter characterizing the relative importance of the attractive potential wells of $\phi^{\rm SE}_{\alpha>1/4}(R)$ and $\phi^{\rm Kohn}(R)$ as $\eta = \left|{\Delta \phi^{\rm SE}_{\alpha>1/4}(R_{\rm min}^{\rm SE})}/{\Delta \phi^{\rm Kohn}(R_{\rm min}^{\rm Kohn})} \right|$,
where $\Delta \phi^{\rm SE}_{\alpha>1/4}(R_{\rm min}^{\rm SE})$ and $\Delta \phi^{\rm Kohn}(R_{\rm min}^{\rm Kohn})$ are, respectively, the values of $\phi^{\rm SE}_{\alpha>1/4}(R)$ and $\phi^{\rm Kohn}(R)$ at the bottoms of their respective potential wells at $R_{\rm min}^{\rm SE}$ and $R_{\rm min}^{\rm Kohn}$, closest to the charge $Q$.

For $\phi^{\rm SE}_{\alpha>1/4}(R)$, we have $R_{\rm min}^{\rm SE} \approx {3\pi}/{4 A_1}$,
\begin{eqnarray}
\Delta \phi^{\rm SE}_{\alpha>1/4}(R_{\rm min}^{\rm SE}) &\approx& \frac{4A_1}{3\pi}\left(b-1\right)\exp\left(-\frac{3\pi}{4}\frac{A_2}{A_1}\right),
\end{eqnarray}
and for $\phi^{\rm Kohn}(R)$, we have $R_{\rm min}^{\rm Kohn} \approx \pi\sqrt{3\alpha_0}$,
\begin{eqnarray}
\Delta \phi^{\rm Kohn}(R_{\rm min}^{\rm Kohn}) &\approx& \frac{\alpha_0}{9\beta^2}\frac{1}{\pi^3\left(3\alpha_0\right)^{3/2}}.
\end{eqnarray}
This gives
\begin{equation}
\eta \approx 36 \pi^2 \sqrt{3\alpha_0}(b-1) A_1 \exp\left(-\frac{3\pi}{4}\frac{A_2}{A_1}\right).  \label{eq:eta}
\end{equation}
The dependence of $\eta$ on $a_B/r_0$ is shown in Fig.~\ref{fig:eta}. It is seen from Figs~\ref{fig:phi_comparison} and \ref{fig:eta} that the depth ratio of the SE attractive potential well and of the first attractive well in the ``Kohn'' potential is less than unity, for all plasma densities including that at which the Shukla-Eliasson's attraction is predicted to be the strongest (at the maximum of $\alpha$).

This suggests that the proper treatment of charge shielding should necessarily be a kinetic one, as the GQHD treatment does not take into account the important (and dominant) part of the attractive force. Another reason that warrants a kinetic treatment of the problem is that the GQHD model's response is only valid for $k\lambda_F\ll 1$~\cite{Vlad_Tysh_UFN_2011}, where $\lambda_F=v_F/\omega_{pe}$ is the Thomas-Fermi length. Yet in Ref.~\cite{SE_PRL_2012} the test charge potential was obtained through integrating its Fourier transform $\phi_k=4\pi Q/k^2\varepsilon(0,k)$ over the entire range of $k$ from $0$ to $\infty$, despite the fact that $\varepsilon(0,k)$ given by their GQHD model is not valid at large $k$. A fully kinetic treatment is free from this limitation, yielding the response that is valid for any $k$. However, a proper way of accounting for exchange interaction and correlations within the framework of kinetic theory is not yet fully developed. Further work in this direction is being done and results will be reported elsewhere.

This work was partially supported by the Australian Research Council and the Ministry of Education of the Russian Federation. The authors thank P.K. Shukla and B. Eliasson for useful and inspiring discussions.


\end{document}